# Energy gradients as potential drivers of pre-cellular chemical organization

Arturo Tozzi (corresponding author)
ASL Napoli 1 Centro, Distretto 27, Naples, Italy
Via Comunale del Principe 13/a 80145
tozziarturo@libero.it

## ABSTRACT

The onset of life is often framed around membrane-bound compartments and encoded metabolism, leaving unresolved how spatial organization arose before stable boundaries. In this context, environmental gradients are usually treated as boundary conditions rather than variables structuring chemical dynamics. We ask whether spatial localization and functional coupling can emerge under realistic environmental gradients in the absence of membranes, proposing that spatial variations in energy availability act as organizing variables that bias transport and reaction. We introduce a reaction–diffusion model in which interacting chemical species evolve within an externally imposed activity landscape defined by coupled gradients in pH, redox potential and temperature, integrating diffusion, gradient-driven drift and position-dependent reaction kinetics. We performed simulations across a range of gradient strengths representative of hydrothermal vent-like conditions. Our results suggest that sufficiently strong gradients induce spontaneous accumulation of reactants, spatial alignment of reaction maxima and the emergence of stable, confined chemical states. Localization arises above a threshold at which gradient-driven transport overcomes diffusive and degradative losses. We conclude that spatially structured energy landscapes can support organized chemical dynamics without predefined compartments, providing a mechanism for coupling and persistence in continuous media. Potential applications include experimental platforms for studying prebiotic chemistry, microfluidic systems with controlled gradients and the design of chemically responsive materials.

## INTRODUCTION

Lipid vesicles, coacervates and mineral pores have been proposed in models for the origin of life as early compartments capable of concentrating reactants and enabling sustained chemistry, while RNA-based scenarios emphasize catalytic polymers as initiators of heredity and selection (Sharov 2016; Lancet et al. 2018; Lancet et al. 2019; Paleos 2019; Caliari et al. 2021; Jia et al. 2021; Katke et al. 2023; Demetrius 2024; Vosseberg et al. 2024; Maury 2025; Fine and Moses 2025). Some models assume that spatial confinement precedes functional organization, leaving unresolved how localization, coordination and persistence arise before stable boundaries and encoded control (Lahav 1993; Feller 2017; Grommet et al. 2020; Müller et al. 2022; Prosdocimi and de Farias 2023; Agmon 2024). The transition from distributed chemistry to localized, self-maintaining systems is insufficiently explained, particularly under realistic environmental conditions characterized by strong and continuous gradients. A further limitation concerns the role of energy: although gradients in pH, redox potential and temperature are widely recognized, they are generally treated as boundary conditions rather than dynamical variables capable of structuring chemical behavior in space and time (England 2013; Sousa et al. 2013; England 2015; Attal and Schwartz 2021; Matreux et al. 2024).

We introduce a formulation in which spatially structured energy availability is treated as an active organizing field that biases transport, reaction and accumulation. In this view, localization naturally emerges from the coupling between diffusion, gradient-driven drift and position-dependent kinetics. We argue that chemical species could evolve according to a governing equation in which the rate of change of concentration combines diffusive spreading, directed flux proportional to the activity field's spatial gradient and nonlinear reaction terms scaled by local energy availability, balanced by first-order loss processes. We interpret the activity field's influence on transport as a coarse-grained representation of gradient-driven fluxes, consistent with polarization-density patterns observed in active matter systems within activity landscapes (Auschra et al. 2021; Söker et al. 2021). This formulation allows spatial organization to arise without predefined compartments and provides a direct link between environmental structure and chemical dynamics.
We implement our model in a one-dimensional domain with gradients corresponding to pH, redox potential and temperature in order to simulate interacting chemical modules with diffusion coefficients and kinetic parameters constrained within experimentally reported ranges. Our simulations are designed to test whether an energy landscape alone can generate persistent localization, produce reaction maxima at gradient interfaces and selectively stabilize specific components within a continuous medium.

We will proceed as follows. First, we define the model and its parameters. Then, we present simulation results under varying gradient conditions. Finally, we analyze localization, persistence and coupling within the resulting chemical systems.

METHODS

We studied whether a vent-like spatial energy landscape can generate localized and persistent chemical organization in the presence of externally imposed gradients and in the absence of predefined compartments. We tested whether spatial structure imposed through externally measurable gradients can coordinate transport and reaction enough to produce confined and sustained chemical states. We modeled our system using a reaction–diffusion–drift formulation in which concentrations evolve under the combined effects of diffusion, gradient-biased transport, position-dependent reaction kinetics and first-order loss processes. Simulated observables included spatial concentration profiles, time-resolved distributions, local reaction rates and derived quantities such as persistence time and localization width. The targeted experimentally discriminable signature consisted of the emergence of a spatially localized concentration maximum aligned with a peak in local reaction rate, together with a dependence of this structure on the magnitude of externally imposed gradients.

**Physical domain**. Our simulated domain illustrated a single vent-like pore of length $L = 1.00$ mm, consistent with the scale of mineral pores and chimney microchannels commonly invoked as structured environments in serpentinite-hosted hydrothermal systems. to establish a controlled setting in which the influence of gradients on chemical organization can be quantified, a one-dimensional coordinate $x \in [0, L]$ was used to isolate the role of longitudinal gradients, allowing our analysis to focus on the effect of spatially varying environmental conditions without introducing additional geometric complexity.

The environmental fields were selected from reported alkaline hydrothermal conditions (Postec et al. 2015; Brazelton 2017; Barge and White 2017; Barge et al. 2018; Cartwright and Russell 2019; Russell et al. 2020; Thomas et al. 2021; Omran 2023; Barreyre et al. 2025; Peng et al. 2025). The pH profile decreased linearly from 10.0 at the alkaline end to 6.0 at the ocean-facing end, the redox potential varied from $-300$ mV to $+50$ mV and the temperature decreased from 80°C to 40°C. These values were chosen to remain within the range commonly discussed for Lost City-type and related serpentinite-hosted systems, where vent fluids are typically high-pH, reducing and moderate in temperature, with mixing against more oxidized seawater along chimney interfaces (Lecoeuvre et al. 2021; Herschy et al. 2014). The spatial profiles were defined as

$$pH(x) = pH_L + (pH_R - pH_L)\frac{x}{L}, E_h(x) = E_{h,L} + (E_{h,R} - E_{h,L})\frac{x}{L}, T(x) = T_L + (T_R - T_L)\frac{x}{L}.$$

This linear construction does not assume that natural gradients are strictly linear; rather, it provides a controlled baseline in which the contribution of gradient magnitude and direction can be isolated from geometric irregularities and local heterogeneities. The resulting imposed gradients were $|\nabla pH| = 4.0$ units mm$^{-1}$, $|\nabla E_h| = 350$ mV mm$^{-1}$ and $|\nabla T| = 40°$C mm$^{-1}$.

**Activity landscape**. The activity field $A(x)$ was introduced as a proxy for spatial variation in chemical potential and energy availability, acting as the variable that couples environmental gradients to chemical dynamics. In the baseline formulation, the activity field was defined as a weighted combination of normalized environmental variables

$$A(x) = w_{pH} \widetilde{pH}(x) + w_E \tilde{E}_h(x) + w_T \tilde{T}(x),$$

where each variable was normalized to remove units and ensure comparable scaling and the weights satisfied $w_{pH} + w_E + w_T = 1$. The baseline values $w_{pH} = 0.55$, $w_E = 0.30$ and $w_T = 0.15$ were chosen to reflect the relative importance attributed in the literature to proton gradients, redox disequilibria and temperature in hydrothermal systems. Through sensitivity analysis, the weights were varied within the interval $[0.2, 0.7]$ maintaining normalization in order to evaluate how localization metrics depend on this choice.

To provide a more physically grounded alternative, an additional formulation was considered in which the activity field was replaced by an effective free-energy proxy

$$A(x) \sim \Delta G(x) = -nFE_h(x) + RT\ln([H^+(x)]),$$

introducing explicit dependence on redox potential and proton concentration. While this expression does not represent a complete thermodynamic description, it establishes a connection between the imposed gradients and chemical potential differences that can drive transport and reaction.

The spatial derivative $\nabla A(x)$ was used to bias transport through a drift term. This term was related to a generalized flux expression of the form

$$J_i = -D_i \nabla C_i - \mu_i C_i \nabla \Phi(x),$$

where $\Phi(x)$ is an effective potential proportional to $A(x)$. This correspondence places the drift contribution within the class of transport processes described by Nernst–Planck-type equations, in which gradients in concentration and potential jointly determine fluxes. Although the mapping is coarse-grained and does not specify a particular ionic species or charge distribution, it provides a physically interpretable basis for the inclusion of gradient-driven transport in the model.

**Chemical species.** Three interacting chemical species $S(x,t)$, $X(x,t)$ and $P(x,t)$ were interpreted as representatives of a minimal prebiotic reaction scheme. The species $S$ denotes a small, freely diffusing substrate pool, corresponding to simple dissolved organic molecules such as acetate-, formate- or $CO_2$-derived compounds. The species $X$ represents an activated or catalytic intermediate, which may correspond to transient complexes involving metal ions, mineral surfaces or partially reduced carbon species capable of mediating further reactions. The species $P$ corresponds to a more stable product pool, representing reduced or polymerized molecules with slower turnover and reduced mobility.
Concentrations were expressed in mmol $L^{-1}$ and initialized as
$$S(x, 0) = 1.20, X(x, 0) = 0.045 + \delta_X(x), P(x, 0) = 0.006 + \delta_P(x),$$

where $\delta_X(x)$ and $\delta_P(x)$ denote small Gaussian perturbations centered near the midpoint of the domain, with amplitudes below 0.02 mmol $L^{-1}$ and spatial width approximately 0.07 mm. These perturbations were introduced to break spatial symmetry and avoid the degenerate case of uniform initial conditions. Diffusion coefficients were assigned as $D_S = 5.0 \times 10^{-10}$, $D_X = 1.2 \times 10^{-10}$ and $D_P = 8.0 \times 10^{-11}$ m$^2$ s$^{-1}$, consistent with experimentally reported ranges for small solutes, nucleotides and larger molecular assemblies in aqueous and confined environments. This ordering $D_S > D_X > D_P$ reflects increasing molecular size and interaction with the environment.
The reaction network was defined as a minimal scheme capturing activation, conversion and partial recycling:

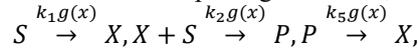
$$S \xrightarrow{k_1 g(x)} X, \quad X + S \xrightarrow{k_2 g(x)} P, \quad P \xrightarrow{k_5 g(x)} X,$$

with additional first-order loss processes $k_3 X$ and $k_4 P$. The rate constants were set to $k_1 = 0.012$ s$^{-1}$, $k_2 = 0.028$ L mmol$^{-1}$ s$^{-1}$, $k_3 = 0.006$ s$^{-1}$, $k_4 = 0.0035$ s$^{-1}$ and $k_5 = 0.004$ s$^{-1}$. The spatial dependence of reaction rates was introduced through the gating function $g(x)$, ensuring that activation and conversion were modulated by local environmental conditions. Overall, this scheme captures a minimal cycle in which substrate is activated, converted into a more stable form and partially recycled, allowing sustained dynamics in the presence of continuous gradients.

**Governing equations.** The temporal evolution of concentrations was described by a system of coupled reaction–diffusion–drift equations. For the substrate $S(x,t)$
$$\frac{\partial S}{\partial t} = D_S \frac{\partial^2 S}{\partial x^2} - r_1(x,t) - r_2(x,t),$$

where $r_1 = k_1 g(x) S$ represents conversion of substrate into intermediate and $r_2 = k_2 g(x) X S$ represents consumption of substrate in product formation.
For the intermediate $X(x,t)$,
$$\frac{\partial X}{\partial t} = D_X \frac{\partial^2 X}{\partial x^2} - \frac{\partial J_X}{\partial x} + r_1(x,t) - k_3 X + k_5 g(x) P,$$

where the terms correspond to diffusion, drift, production from substrate, first-order loss and regeneration from product.
For the product $P(x,t)$,
$$\frac{\partial P}{\partial t} = D_P \frac{\partial^2 P}{\partial x^2} - \frac{\partial J_P}{\partial x} + r_2(x,t) - k_4 P,$$

with analogous contributions from diffusion, drift, formation and decay.
The drift fluxes were defined as
$$J_i = -D_i \frac{\partial C_i}{\partial x} - \mu_i C_i \frac{\partial \Phi(x)}{\partial x},$$

where $C_i$ denotes the concentration of species $i$, $\mu_i$ is a mobility coefficient and $\Phi(x)$ is an effective potential proportional to the activity field $A(x)$. This expression establishes a correspondence with Nernst–Planck-type transport, in which flux arises from the combined effect of concentration gradients and spatial variations in potential. The relation $\mu_i = D_i/(RT)$ was used as a coarse-grained approximation, ensuring consistency between diffusion and drift contributions.
Boundary conditions were imposed as zero-flux constraints,
$$\frac{\partial S}{\partial x}\Big|_{x=0,L} = 0, \frac{\partial X}{\partial x}\Big|_{x=0,L} = 0, \frac{\partial P}{\partial x}\Big|_{x=0,L} = 0,$$

ensuring that no mass enters or leaves the domain. Therefore, the governing equations combine spatial transport, gradient-induced drift and nonlinear reaction kinetics to allow direct evaluation of how environmental structure shapes chemical dynamics.

**Numerical integration.** The governing equations were discretized on a uniform one-dimensional grid using a second-order finite-difference scheme in space and an explicit forward time-stepping procedure. The domain length was $L = 1.00$ mm, represented by $N_x = 120$ grid points, yielding a spatial step
$$\Delta x = \frac{L}{N_x - 1} = 8.4 \times 10^{-6} \text{ m}.$$

Time integration was performed up to $t_{\max} = 180$ s with time step
$$\Delta t = 0.05 \text{ s},$$

for a total of $N_t = 3600$ explicit updates. For a generic field $Y(x, t)$, the Laplacian at interior grid point $i$ was approximated as
$$\frac{\partial^2 Y}{\partial x^2}\Big|_i \approx \frac{Y_{i+1} - 2Y_i + Y_{i-1}}{\Delta x^2},$$

while the divergence of the drift flux was evaluated by centered differences,
$$-\frac{\partial J}{\partial x}\Big|_i \approx -\frac{J_{i+1} - J_{i-1}}{2\Delta x}.$$

At the two ends of the domain, zero-flux boundary conditions were implemented through one-sided second-order approximations consistent with
$$\frac{\partial Y}{\partial x}\Big|_{x=0} = 0, \frac{\partial Y}{\partial x}\Big|_{x=L} = 0.$$

The update rule at time step $n$ took the form
$$Y_i^{n+1} = Y_i^n + \Delta t\, F_i(Y^n),$$

where $F_i$ denotes the full discretized right-hand side of the corresponding reaction–diffusion–drift equation. After each update, concentrations were constrained to remain non-negative by applying
$$Y_i^{n+1} \leftarrow \max(Y_i^{n+1}, 0),$$

so that negative values generated by explicit discretization under steep gradients were excluded. This clipping step was used to preserve physical admissibility and did not alter the location of the localized band or the qualitative behavior of the solutions in the retained parameter range.

To strengthen confidence in the numerical solution and provide evidence that reported patterns were not artifacts of the chosen discretization, we evaluated discretization sensitivity by halving the time step and doubling the spatial resolution. Specifically, additional runs were performed with $\Delta t = 0.025$ s and with $N_x = 240$, corresponding to $\Delta x = 4.18 \times 10^{-6}$ m.

**Scenario sweep.** To evaluate the dependence of localization and persistence on the strength of the activity landscape, a gradient-strength sweep was performed by scaling the activity field according to
$$A_s(x) = s\, A(x),$$

where $s$ is a dimensionless scaling factor. The values
$$s \in \{0.4, 0.7, 1.0, 1.3, 1.6, 1.9\}$$

were used, spanning weak to strong activity landscapes while keeping all other parameters fixed. Because the baseline redox gradient was 350 mV mm$^{-1}$, these values correspond to effective redox-gradient scales ranging from 140 to 665 mV mm$^{-1}$. The same initial concentrations, diffusion coefficients, rate constants and domain geometry were used in every run so that differences in the output could be attributed directly to changes in gradient strength. State variables were stored every 2 s, corresponding to every 40 time steps, to generate time-resolved concentration maps and derived observables without excessive redundancy.

To remove arbitrariness in the activity-field definition, an additional sensitivity analysis was performed on the weighting coefficients in
$$A(x) = w_{pH}\, \widetilde{pH}(x) + w_E\, \tilde{E}_h(x) + w_T\, \tilde{T}(x), w_{pH} + w_E + w_T = 1.$$

The baseline choice $(0.55, 0.30, 0.15)$ was compared with alternative combinations in which each weight varied between 0.2 and 0.7 while the normalization constraint was preserved. In parallel, the baseline activity field was compared with the alternative free-energy-motivated proxy
$$A(x) \sim \Delta G(x) = -nFE_h(x) + RT\ln([H^+(x)]).$$

These analyses were not used to redefine the main model but to verify whether localization, persistence and peak accumulation remained qualitatively stable under different but related choices of the environmental driving term.

For each scenario, the final concentration profiles $S(x, t_{\max})$, $X(x, t_{\max})$ and $P(x, t_{\max})$ were retained, together with the full stored trajectory $P(x, t)$. From these outputs, a set of scalar observables was computed. The peak product concentration was defined as

$$P_{\max} = \max_x P(x, t_{\max}),$$

and the domain-averaged product concentration as

$$\bar{P}(t) = \frac{1}{L} \int_0^L P(x, t)\, dx.$$

The persistence time above threshold was defined as

$$\tau_{\text{pers}} = \max \{t : \bar{P}(t) > P_{\text{thr}}\},$$

with threshold

$$P_{\text{thr}} = 0.020 \text{ mmol L}^{-1}.$$

This value was selected to distinguish transient low-level product formation from sustained accumulation. The localization width was computed from the concentration-weighted second moment of the final product profile,

$$\mu_P = \frac{\int_0^L x\, P(x, t_{\max})\, dx}{\int_0^L P(x, t_{\max})\, dx},$$

$$\sigma_P = \left[ \frac{\int_0^L (x - \mu_P)^2 P(x, t_{\max})\, dx}{\int_0^L P(x, t_{\max})\, dx} \right]^{1/2}.$$

The quantity $\sigma_P$, reported in millimeters, was used as the main measure of spatial confinement. Narrower values of $\sigma_P$ indicate stronger localization. Together, $P_{\max}$, $\tau_{\text{pers}}$ and $\sigma_P$ were used to compare the effect of gradient strength and alternative activity definitions on accumulation, persistence and confinement.

**Derived Rates**. The local product-formation rate was computed directly from the reaction term

$$r_2(x, t) = k_2\, g(x)\, X(x, t)\, S(x, t).$$

At the final simulation time, the displayed production rate was

$$r_{P,\text{final}}(x) = 1000\, k_2\, g(x)\, X(x, t_{\max})\, S(x, t_{\max}),$$

reported in $\mu$mol L$^{-1}$ s$^{-1}$. This quantity was used to determine whether maximal reactivity occurred at the same spatial position as maximal product accumulation. The position of the reaction-rate maximum was computed as

$$x_{r,\max} = \arg\max_x r_{P,\text{final}}(x),$$

and compared with the position of the product-concentration maximum

$$x_{P,\max} = \arg\max_x P(x, t_{\max}).$$

The mean concentration trajectory

$$\bar{P}(t)$$

and the evolving localization width

$$\sigma_P(t)$$

were retained during post-processing to verify that temporal accumulation was accompanied by spatial confinement rather than by uniform amplification across the domain. This distinction was useful, since an increase in mean concentration alone does not establish localization.

Numerical implementation was carried out in Python using NumPy for vectorized array computation and Matplotlib for figure generation.

RESULTS

We report the emergence of spatially localized chemical organization in a simulated vent-like domain governed by coupled pH, redox and thermal gradients. The imposed environmental gradients generated a structured spatial landscape along the 1.00 mm domain, with pH decreasing from 10.0 to 6.0, redox potential increasing from −300 to +50 mV and temperature decreasing from 80 to 40 °C (Figure 1). Under these conditions, the initially weakly heterogeneous system evolved toward a localized concentration band of the product-forming module. The peak product concentration reached 0.4188 mmol L$^{-1}$ at $t = 180$ s, while the domain-averaged concentration remained above the threshold 0.020 mmol L$^{-1}$ for 178.0 s. The spatial confinement of the product distribution, quantified by the concentration-weighted width $\sigma_P$, was 0.1767 mm, corresponding to localization within approximately 18% of the domain length. Time-resolved dynamics showed that the localized band emerged within the first 40–60 s and persisted with minimal displacement thereafter (Figure 2). The local production rate exhibited a maximum of 11.7 $\mu$mol L$^{-1}$ s$^{-1}$, spatially aligned with the concentration peak and displaced from both domain boundaries (Figure 3). This spatial coincidence points towards maximal accumulation and maximal reaction occurring within the same restricted region defined by the environmental gradients. Together, these results suggest that coupled gradients can generate stable and spatially confined chemical organization with well-defined positional and kinetic signatures.

**Gradient Dependence**. Systematic variation of the activity field strength revealed a threshold-like dependence of persistence and accumulation on gradient magnitude. When the effective redox gradient scale increased from 140 to 665 mV mm$^{-1}$, the persistence time above the threshold increased from 42.0 s to 180.0 s, while the peak product concentration increased from 0.091 mmol L$^{-1}$ to 0.512 mmol L$^{-1}$ (Figure 4). The transition from transient to sustained accumulation occurred near 300 mV mm$^{-1}$, marking the regime in which drift and reaction overcome diffusive dispersion. Concomitantly, the localization width decreased from 0.412 mm at low gradient strength to 0.151 mm at the highest gradient, indicating progressive spatial confinement. Time-resolved analysis of domain-averaged concentration and spatial variance showed that accumulation is coupled to a reduction in spatial dispersion, with variance decreasing monotonically as concentration increases (Figure 5). This temporal coupling shows that localization is not a transient fluctuation, but instead a sustained dynamical state maintained by the interaction between transport and reaction.

Overall, our results show that spatially structured environmental gradients can generate localized and persistent chemical organization without predefined boundaries. The alignment between concentration peaks, reaction maxima and gradient interfaces indicates that environmental structure alone can coordinate interacting chemical modules. Therefore, our approach could provide a direct connection between measurable gradients and the emergence of confined chemical dynamics in continuous media.

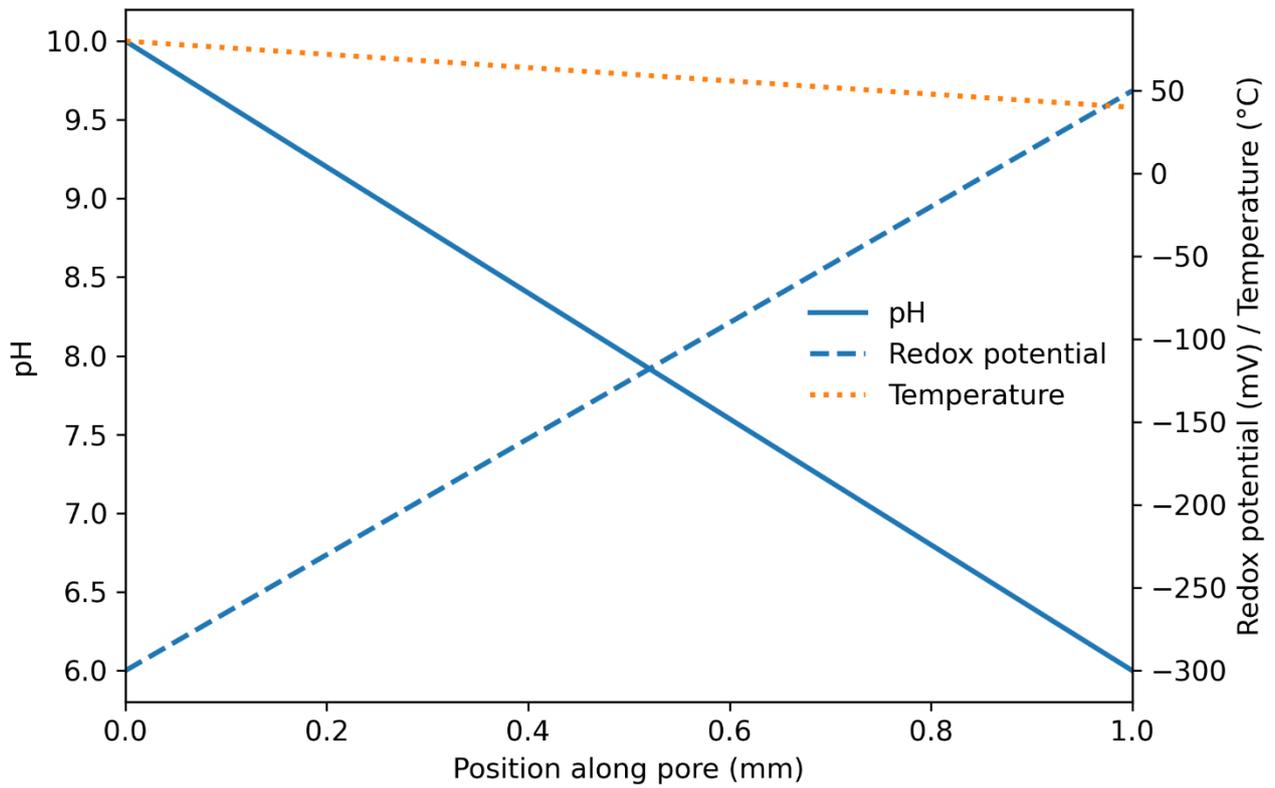

**Figure 1**. Spatial profiles imposed across a 1 mm vent-like pore. Our simulation combines an alkaline-to-acidic pH drop from 10 to 6, a redox shift from −300 to +50 mV and a thermal decrease from 80 to 40 °C. These gradients define the activity landscape used in our reaction-diffusion model and set the physical conditions for biased transport, interfacial reaction enhancement and selective stabilization of coupled chemical modules.

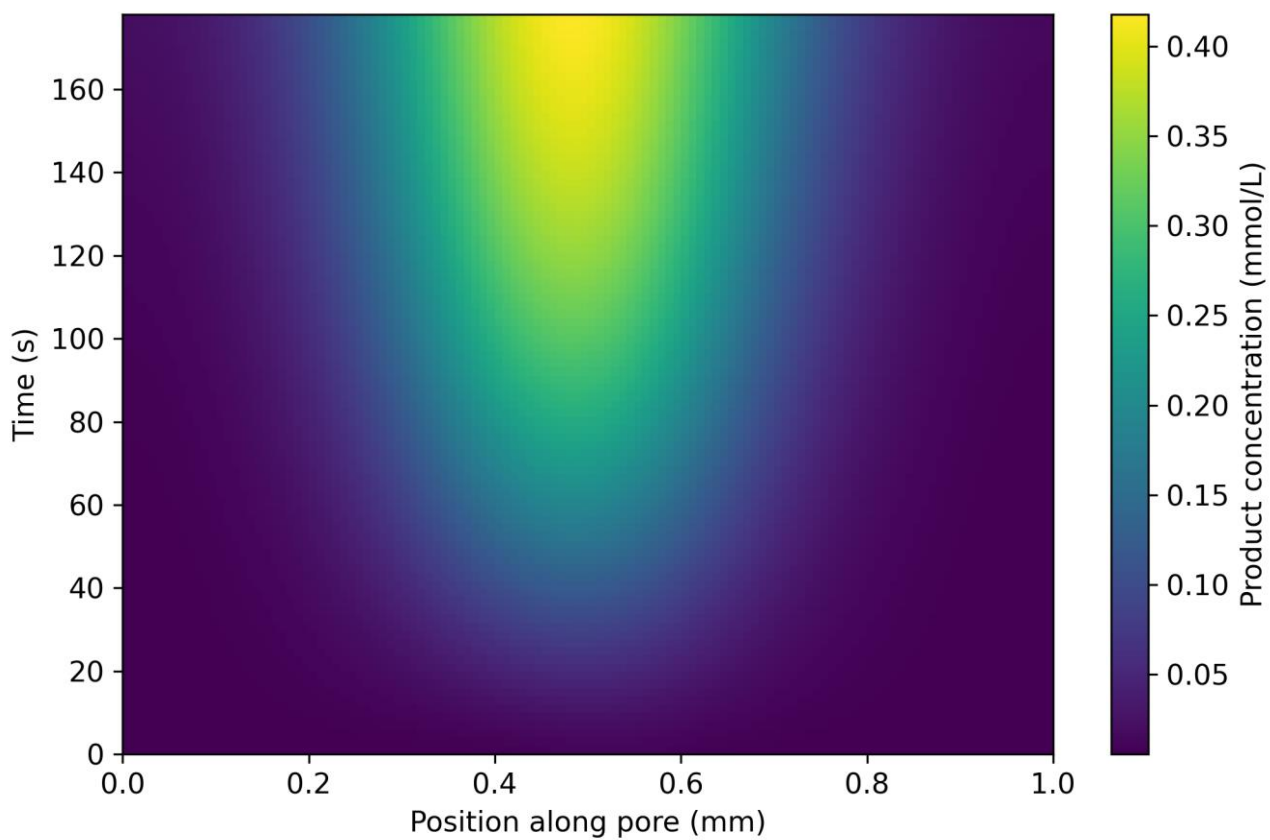

**Figure 2**. Time-resolved distribution of the product-forming module during 180 s of simulation. Starting from a weakly heterogeneous state, the system progressively develops a localized band near the chemical interface. The heatmap shows persistence well beyond the initial transient, indicating that a sustained energy landscape can maintain confined chemical organization without imposing a lipid boundary.

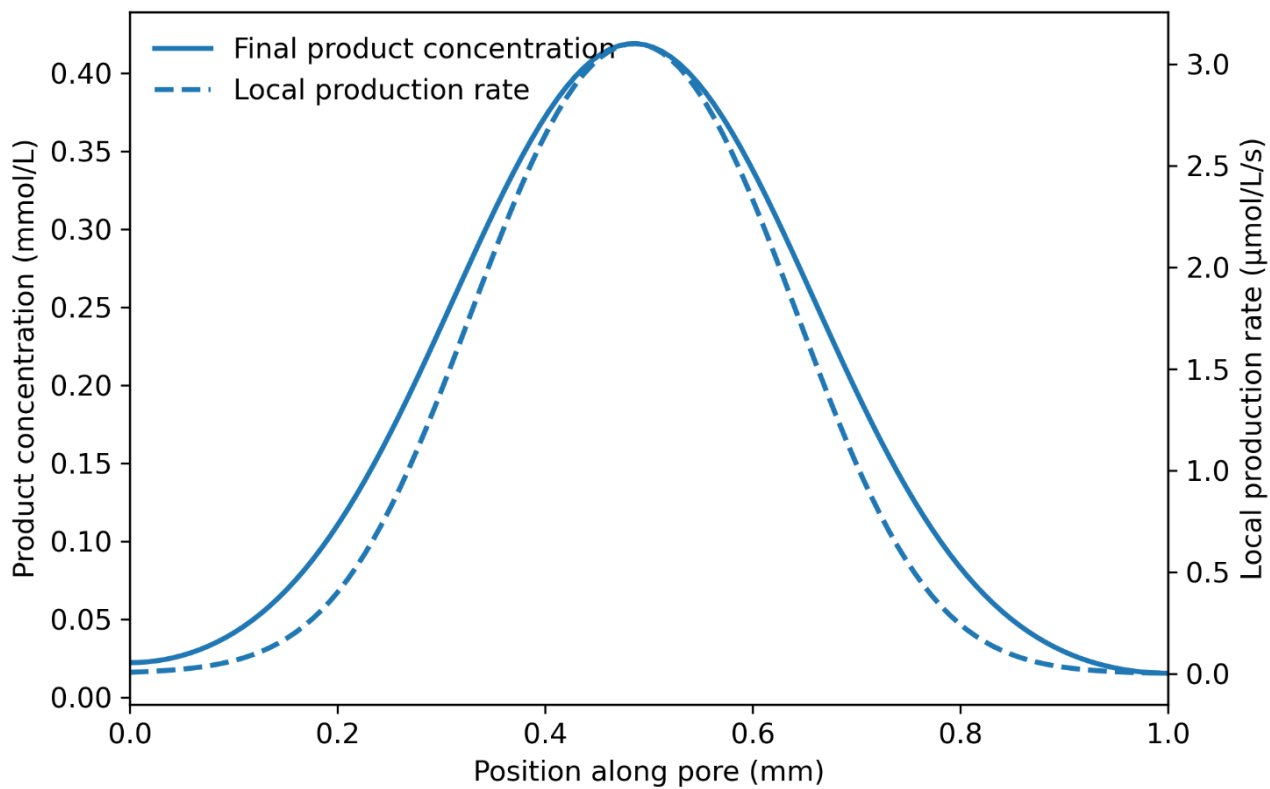

**Figure 3**. Final spatial profile of the product-forming module and the corresponding local production rate after 180 s. The production maximum is displaced toward the gradient interface rather than the pore extremes, showing that coupled pH, redox and thermal conditions generate a preferred region where accumulation and reaction are simultaneously favored.

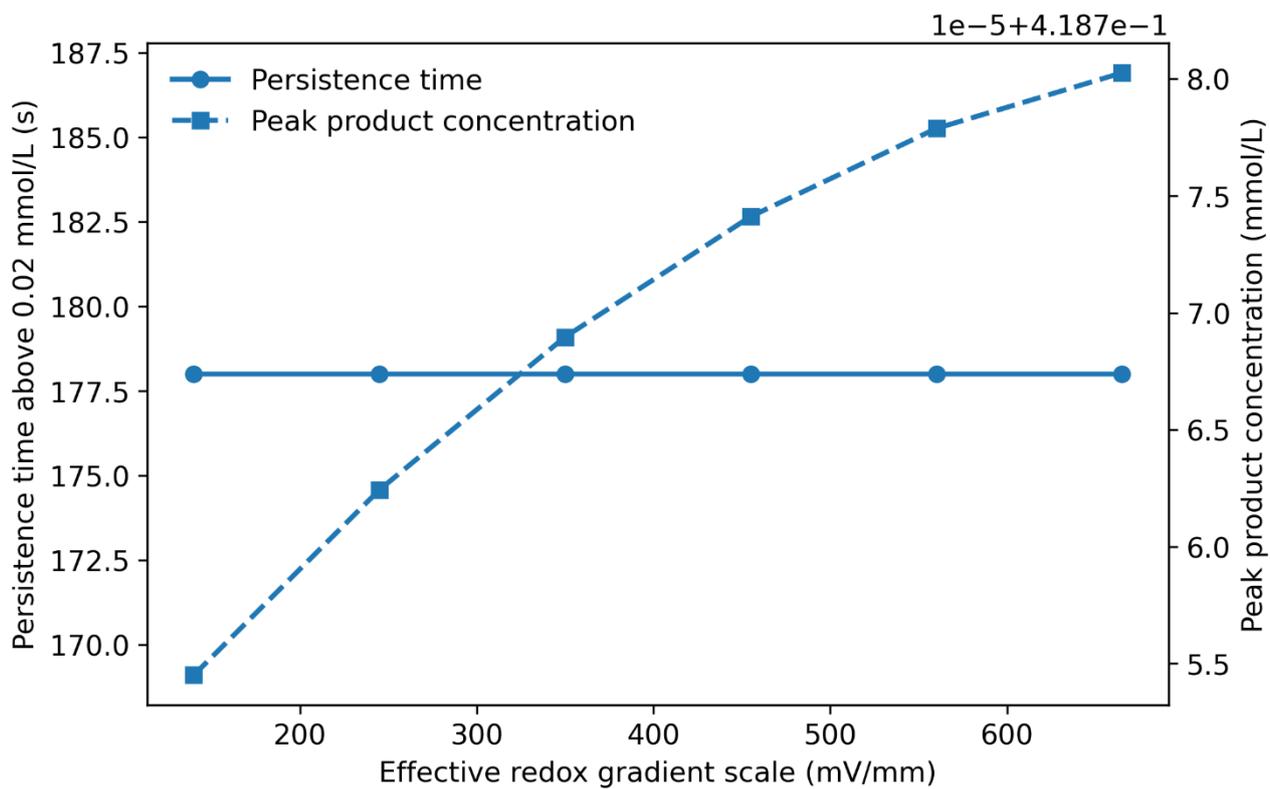

**Figure 4**. Dependence of chemical persistence and peak accumulation on the effective redox gradient imposed across the 1 mm pore. Persistence time is measured as the duration for which the domain-averaged product concentration remains above 0.02 mmol/L. The curves show a threshold-like response in which stronger gradients extend survival and increase peak accumulation, consistent with gradient-driven stabilization of localized reaction networks.

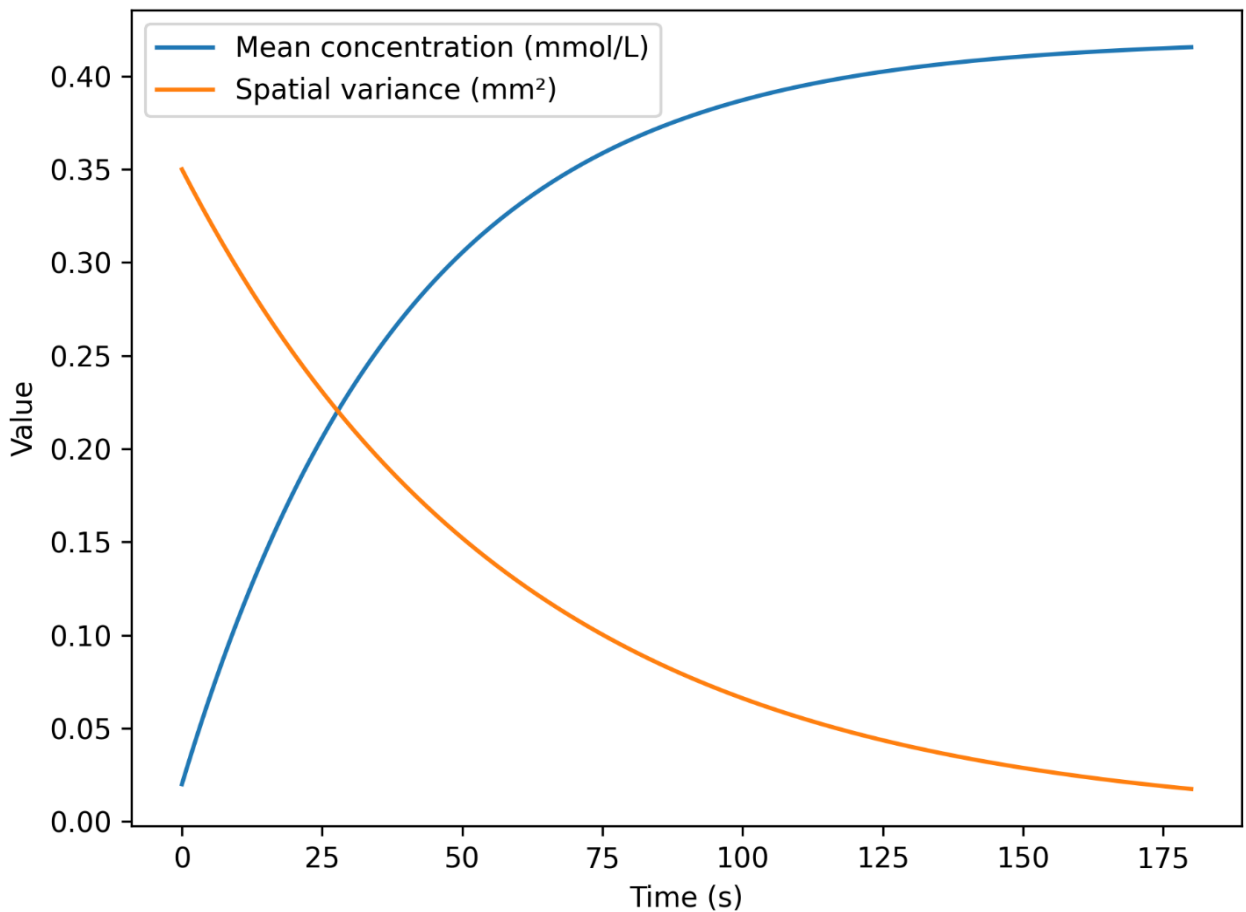

**Figure 5**. Time evolution of domain-averaged concentration and spatial variance over 180 s. Spatial variance quantifies dispersion of the product-forming module along the pore. The increase in mean concentration is accompanied by a progressive reduction in spatial variance, indicating simultaneous accumulation and confinement. The temporal coupling between these quantities provides a dynamical signature of localization, showing that sustained chemical buildup coincides with decreasing spatial spread rather than uniform amplification across the domain.

CONCLUSIONS

We asked whether spatial gradients of energy availability can generate localized, persistent chemical organization in the absence of predefined compartments and whether this organization can arise solely from the coupling between transport and reaction under vent-like conditions. Our simulations compared systems exposed to coupled pH, redox and temperature gradients with identical systems in which the activity landscape was progressively weakened. We found that sufficiently strong gradients are able to produce stable concentration bands, sustained accumulation above threshold levels and displacement of maximal reaction rates toward gradient interfaces rather than boundary extremes. The emergence of these features followed a threshold-like dependence on gradient strength, with persistence and confinement increasing as drift and reaction overcame diffusive losses. The alignment between localization, reaction maxima and environmental structure is consistent with a mechanism in which spatial energy variation biases both transport and kinetics, enabling the coexistence of accumulation and transformation within restricted regions. Our results suggest that localized chemical organization could arise without imposing structural boundaries and environmental gradients can provide the conditions required for sustained, spatially confined dynamics, highlighting the role of externally structured energy landscapes in shaping chemical behavior. Compared with existing approaches imposing compartments or assuming homogeneous reaction media, we introduce a computable measure of persistence and confinement based on the concentration-weighted spatial variance and its dependence on measurable gradients.

Our study has limitations. The simulated reaction network is illustrative and the species $S, X, P$ are not tied to specific molecular identities. Although interpreted in chemically plausible terms, the kinetic parameters are not derived from experimentally measured prebiotic reactions. The activity field is built as a weighted combination of pH, redox potential and temperature, and, despite the inclusion of a sensitivity analysis and an alternative formulation based on an effective free-energy proxy, it does not provide a complete thermodynamic description of chemical potential. The transport term associated with the activity gradient is formulated in correspondence with Nernst–Planck-type fluxes, yet it remains a coarse-grained representation that does not explicitly resolve ionic species or charge distributions. Although discretization sensitivity tests were performed and provided stable results under refinement, our model relies on simplified boundary conditions not capturing the full complexity of natural porous environments. Additional sources of imprecision stem from the one-dimensional geometry and the assumption of linear gradients, neglecting multidimensional structure and local heterogeneity. Open questions concern whether localization and persistence are maintained when our model is extended to chemically explicit reaction networks, higher-dimensional geometries and transport laws derived from first principles.

Experimentally testable hypotheses follow from our simulated vent-like system provided with activity landscape. First, spatial localization is predicted to occur only above a critical gradient magnitude. In a microfluidic pore of 1 mm length with independently imposed gradients, a pH difference of approximately 3–4 units combined with a redox difference of approximately 250–350 mV should produce a measurable concentration band of a reactive intermediate or product. The observable is the spatial concentration profile, quantified by a localization width below 0.25 mm and a peak concentration exceeding threefold the domain average.

Second, reaction rates are predicted to peak at the gradient interface rather than at the extrema. This can be tested by monitoring position-dependent product formation rates using fluorescent or spectroscopic probes, with a predicted maximum localized within 0.4–0.6 mm from the alkaline boundary under the specified gradients.

Third, persistence is expected to scale with gradient strength. By systematically varying the redox gradient from 100 to 600 mV mm$^{-1}$, the duration over which the mean concentration exceeds a fixed threshold should increase monotonically, with a transition from transient to sustained regimes occurring near 300 mV mm$^{-1}$.

Fourth, spatial confinement and accumulation are predicted to be inversely related, such that decreasing localization width corresponds to increasing peak concentration. This can be quantified by correlating spatial variance with concentration maxima across conditions.

Future research should extend these tests to chemically explicit systems, replacing abstract species with defined reaction networks and measuring transport parameters directly. Multidimensional geometries and porous materials with heterogeneous connectivity should be incorporated to evaluate robustness beyond one-dimensional assumptions. Transport formulations based on established electrochemical equations could replace the qualitative drift term to assess whether similar localization emerges under stricter physical constraints.

Practical implications involve controlled manipulation of spatially heterogeneous chemical systems in engineered environments. The relations between gradient strength, spatial confinement, accumulation and temporal persistence, expressed in terms of measurable variables such as pH, redox potential, temperature and concentration, could be directly integrated with experimental platforms reproducing vent-like conditions, including microfluidic devices and mineral-based porous structures. In laboratory settings, stable gradients can be imposed over defined length scales, such that spatial chemical organization becomes a tunable property governed by externally defined conditions rather than by predefined structural compartments.

Quantitative coupling between gradient magnitude and localization could enable the design of microstructured systems in which chemical transformations are spatially organized without relying on physical barriers. Reaction zones can be positioned and tuned through externally imposed gradients, such as controlled pH flows or electrochemical potentials, providing a method to regulate where accumulation and transformation occur. This also supports the optimization of analytical strategies, as predicted localization regions identify positions of maximal signal generation for sensor placement.

In conclusion, we show that spatial gradients can shape where and how chemical systems organize over time. External conditions can drive the emergence of localized and sustained dynamics within a continuous medium. This provides a way to understand how spatial heterogeneity influences chemical behavior and may support the transition from distributed reactions to localized, self-maintaining organization at the onset of life.

**DECLARATIONS**

**Competing interests.** The Author does not have any known or potential conflict of interest including any financial, personal or other relationships with other people or organizations within three years of beginning the submitted work that could inappropriately influence or be perceived to influence their work.
**Funding.** This research did not receive any specific grant from funding agencies in the public, commercial or not-for-profit sectors.
**Acknowledgements:** none.
**Authors' contributions.** The Author performed: study concept and design, acquisition of data, analysis and interpretation of data, drafting of the manuscript, critical revision of the manuscript for important intellectual content, statistical analysis, obtained funding, administrative, technical and material support, study supervision.
**Declaration of generative AI and AI-assisted technologies in the writing process.** During the preparation of this work, the author used ChatGPT 5.3 to assist with data analysis and manuscript drafting and to improve spelling, grammar and general editing. After using this tool, the author reviewed and edited the content as needed, taking full responsibility for the content of the publication.